\documentclass[12pt]{article}

\usepackage{amsmath,amssymb,graphicx} 
\usepackage{epsf}
\usepackage{pstricks}
\usepackage{cite}

\newcommand{\beq}{\begin{eqnarray}}
\newcommand{\eeq}{\end{eqnarray}}

\newcommand{\centeron}[2]{{\setbox0=\hbox{#1}\setbox1=\hbox{#2}\ifdim

\wd1>\wd0\kern.5\wd1\kern-.5\wd0\fi
\copy0

\kern-.5\wd0\kern-.5\wd1\copy1\ifdim\wd0>\wd1
                                       \kern.5\wd0\kern-.5\wd1\fi}}
\newcommand{\ltap}{\>\centeron{\raise.35ex\hbox{$<$}}
                               {\lower.65ex\hbox{$\sim$}}\>}
\newcommand{\gtap}{\>\centeron{\raise.35ex\hbox{$>$}}
                               {\lower.65ex\hbox{$\sim$}}\>}

\newcommand\ZZ{\hbox{\zfont Z\kern-.4emZ}}
\font\zfont = cmss10 

\newcommand{\vo}{v^{(0)}}
\newcommand{\ao}{a^{(0)}}
\newcommand{\aone}{a^{(1)}}
\newcommand{\vone}{v^{(1)}}

\begin{document}
\begin{titlepage}
\begin{flushright}
\vphantom{{\tt hep-ph/yymmnnn}}
\end{flushright}

\vskip.5cm
\begin{center}
{\huge \bf The $S$-parameter in\\[.2cm] Holographic Technicolor Models}
\vskip.2cm
\end{center}

\begin{center}
{\bf Kaustubh Agashe$^a$, Csaba Cs\'aki$^b$, Christophe Grojean$^{c,d}$, and Matthew Reece$^b$} \\
\end{center}
\vskip 8pt

\begin{center}
$^a$ {\it Department of Physics, Syracuse University, Syracuse, NY
13244, USA} \\

\vspace*{0.2cm}

$^{b}$ {\it Institute for High Energy Phenomenology\\
Newman Laboratory of Elementary Particle Physics, \\
Cornell University, Ithaca, NY 14853, USA } \\

\vspace*{0.2cm}

$^{d}$ {\it CERN, Theory Division, CH 1211, Geneva 23, Switzerland} \\

\vspace*{0.2cm}

$^{c}$ {\it Service de Physique Th\'eorique, CEA Saclay, F91191, Gif-sur-Yvette, France} \\

\vspace*{0.3cm} {\tt kagashe@phy.syr.edu,
csaki@lepp.cornell.edu, \\christophe.grojean@cern.ch,
mreece@lepp.cornell.edu}
\end{center}

\vglue 0.3truecm

\begin{abstract}

We study the $S$ parameter, considering especially
its sign, in models of
electroweak symmetry breaking (EWSB) in extra dimensions,
with fermions localized near the UV brane.
Such models are conjectured to be
dual to $4D$ strong dynamics triggering EWSB.
The motivation for such a study is that
a negative value of $S$ can significantly ameliorate the
constraints from electroweak precision data on these models,
allowing lower mass scales (TeV or below) for
the new particles and leading to
easier discovery at the LHC.
We first extend an earlier proof of $S>0$ for EWSB
by boundary conditions in arbitrary metric to the case of
general kinetic functions for the gauge fields or arbitrary kinetic mixing.
We then consider
EWSB in the bulk by a Higgs VEV
showing that $S$ is positive for arbitrary metric and
Higgs profile, assuming that the effects
from higher-dimensional operators
in the $5D$ theory are sub-leading and can therefore be neglected.
For the specific case of AdS$_5$ with a power law Higgs profile,
we also show that $S \sim + O(1)$, including effects
of possible kinetic mixing from higher-dimensional operator
(of NDA size)
in the
$5D$ theory.
%
%
Therefore, our
work strongly suggests that $S$ is positive in
calculable models in extra dimensions.

\vskip 3pt \noindent {\bf }
\end{abstract}

\end{titlepage}

\newpage

\section{Introduction}
\setcounter{equation}{0}
\setcounter{footnote}{0}

One of the outstanding problems in particle physics is to understand the mechanism
of electroweak symmetry breaking. Broadly speaking, models of natural electroweak
symmetry breaking rely either on supersymmetry or on new strong dynamics at some
scale near the electroweak scale. However, it has long been appreciated
that if the new strong dynamics is QCD-like,
it is in conflict with precision tests of electroweak observables~\cite{techniEW}. Of particular
concern is the $S$ parameter. It does not violate custodial symmetry; rather, it is directly
sensitive to the breaking of SU(2). As such, it is difficult to construct models that have $S$
consistent with data, without fine-tuning.

The search for a technicolor model consistent with data, then, must turn to non-QCD-like
dynamics. An example is ``walking"~\cite{Walking}, that is, approximately conformal dynamics,
which can arise in theories with extra flavors. It has been argued that such nearly-conformal
dynamics can give rise to a suppressed or even negative contribution to the $S$
parameter~\cite{WalkingS}. However, lacking nonperturbative calculational tools, it is difficult
to estimate $S$ in a given technicolor theory.

In recent years, a different avenue of studying dynamical EWSB models
has opened up via the realization that extra dimensional models~\cite{RS} may provide
a weakly coupled dual description to technicolor type theories~\cite{Arkani-Hamed:2000ds}.
The most studied
of these higgsless models~\cite{Higgsless} is based on an AdS$_5$ background
in which the Higgs is localized on the TeV brane and has a very large VEV, effectively decoupling
from the physics.  Unitarization is accomplished by gauge KK modes, but this leads to a
tension: these KK modes cannot be too heavy or perturbative unitarity is lost, but if they
are too light then there are difficulties with electroweak precision: in particular, $S$ is
large and positive~\cite{BPR}. In this argument the fermions are assumed to be
elementary in
the $4D$ picture (dual to them
being localized on the Planck brane). A possible way out is to assume that the
direct contribution of the EWSB dynamics to the $S$-parameter are compensated by
contributions to the
fermion-gauge boson vertices~\cite{Disguising,Cacciapaglia:2006pk}.
In particular, there exists a scenario where the fermions
are partially composite in which $S \approx 0$~\cite{CuringIlls}, corresponding to
almost flat wave functions for the fermions along the extra dimension.
The price of this cancellation is a percent level tuning in the Lagrangian parameter
determining the shape of the fermion wave functions.
Aside from the tuning itself, this is also undesirable
because it gives the model-builder very little freedom in addressing flavor problems: the fermion
profiles are almost completely fixed by consistency with electroweak precision.

While Higgsless models are the closest extra-dimensional models to traditional technicolor
models, models with a light Higgs in the spectrum do not require light gauge KK modes for
unitarization and can be thought of as composite Higgs models. Particularly appealing
are those where the Higgs is a pseudo-Nambu-Goldstone boson~\cite{CompositeHiggs, SILH}.
In these models, the electroweak constraints are less strong, simply because most of the
new particles are heavy. They still have a positive $S$, but it can be small enough to be
consistent with data. Unlike the Higgsless models where one is forced to delocalize
the fermions, in these models with a higher scale the fermions can be peaked near
the UV brane so that flavor issues can be addressed.

Recently, an interesting alternative direction to eliminating the $S$-parameter constraint
has been proposed in~\cite{HirnSanz}. There it was argued, that by considering holographic models of
EWSB in more general backgrounds with non-trivial profiles of a bulk Higgs field one could
achieve $S<0$. The aim of this paper is to investigate the feasibility of this proposal.
We will focus on the direct contribution of the strong dynamics to $S$. In particular, we imagine that the SM
fermions can be almost completely elementary in the
$4D$ dual picture, corresponding to them being localized near the UV brane. In this case,
a negative $S$ would offer appealing new prospects for model-building
since such values of $S$ are less constrained by data than a positive
value~\cite{Yao:2006px}.
Unfortunately we find that the $S>0$ quite generally, and that
backgrounds giving negative $S$ appear to be pathological.

The outline of the paper is as follows. We first present a general
plausibility argument based purely on 4D considerations that one is unlikely to find
models where $S<0$. This argument is independent from the rest of the paper, and
the readers
interested in the holographic considerations may skip directly to section \ref{review}.
Here we first review the formalism
to calculate the $S$ parameter in quite general models of EWSB
using an extra dimension.
We also extend
%
%
the proof of $S>0$ for BC breaking \cite{BPR} in arbitrary metric
to the case of arbitrary {\em kinetic functions} or localized kinetic mixing terms.
These proofs quite clearly show that no form of boundary condition breaking
will result in $S<0$. However, one may hope that (as argued in~\cite{HirnSanz}) one can
significantly modify this result by using a bulk Higgs with a profile peaked towards the IR
brane to break the electroweak symmetry.  Thus, in the crucial section \ref{bulkbreaking}, we
show that $S>0$ for models with bulk breaking from a scalar VEV as well. Since the gauge
boson mass is the lowest dimensional operator sensitive to EWSB one would expect that this is
already sufficient to cover all interesting possibilities. However, since the Higgs VEV can be
very strongly peaked, one may wonder if other (higher dimensional) operators could become
important as well. In particular, the kinetic mixing operator of $L,R$ after Higgs VEV insertion
would be a direct contribution to $S$. To study the effect of this operator
in section \ref{equivalence},
it is shown that the bulk mass term for axial field can be
converted
to kinetic functions as well,
%
%
making a unified treatment of the effects of bulk mass terms and
the effects of the kinetic
mixing from the higher-dimensional
operator possible.
Although we do not have a general proof that $S > 0$ including
the effects of the bulk kinetic mixing for a general metric
and Higgs profile,
in section \ref{scan} we present a detailed scan
for AdS metric and for power-law Higgs vev profile using the
technique of the previous section for arbitrary kinetic mixings.
We find $S > 0$ once we require that the higher-dimensional
operator is of
NDA size, and that the theory
is ghost-free. We summarize and conclude in section \ref{end}.

\section{A plausibility argument for $S > 0$}
\label{plausibility}
\setcounter{equation}{0}
\setcounter{footnote}{0}

In this section we define $S$ and sketch a brief argument for its
positivity in a general technicolor model. The reader mainly
interested in the extra-dimensional constructions  can skip this
section  since it is independent from the rest of the paper.
However, we think it is worthwhile to try to understand why one
might expect $S > 0$ on simple physical grounds. The only
assumptions we will make are that we have some strongly coupled
theory that spontaneously breaks SU(2)$_L \times $SU(2)$_R$ down to
SU(2)$_V$, and that at high energies the symmetry is restored. With
these assumptions, $S > 0$ is plausible. $S < 0$ would require more
complicated dynamics, and might well be impossible,  though we
cannot prove it.\footnote{For a related discussion of the
calculation of $S$ in strongly coupled theories,
see~\cite{schrock}.}

Consider a strongly-interacting theory with SU(2) vector current $V^a_\mu$ and SU(2) axial vector
current $A^a_\mu$. We define (where $J$ represents $V$ or $A$):
\beq
i \int d^4 x~e^{-i q \cdot x} \left<J^a_\mu(x) J^b_\mu(0)\right> = \delta^{ab} \left(q_\mu q_\nu - g_{\mu \nu} q^2\right) \Pi_J(q^2).
\eeq
We further define the left-right correlator, denoted simply $\Pi(q^2)$, as $\Pi_V(q^2) - \Pi_A(q^2)$.
In the usual way, $\Pi_V$ and $\Pi_A$ are related to positive spectral functions $\rho_V(s)$
and $\rho_A(s)$. Namely, the $\Pi$ functions are analytic functions of $q^2$ everywhere in
the complex plane except for Minkowskian momenta, where poles and branch points can appear
corresponding to physical particles and multi-particle thresholds. The discontinuity across the singularities on the $q^2 > 0$ axis is given by a spectral function.  In particular, there is a dispersion relation
\beq
\Pi_V(q^2) = \frac{1}{\pi} \int_0^{\infty} ds~\frac{\rho_V(s)}{s - q^2 + i \epsilon},
\eeq
with $\rho_V(s) > 0$, and similarly for $\Pi_A$.

Chiral symmetry breaking establishes that $\rho_A(s)$ contains a term
$\pi f_\pi^2 \delta(s)$. This is the massless particle pole corresponding to the Goldstone of the spontaneously broken SU(2) axial flavor symmetry. (The corresponding pions, of course, are eaten once we couple the theory to the Standard Model, becoming the longitudinal components of the $W^\pm$ and $Z$ bosons. However, for now we consider the technicolor sector decoupled from the Standard Model.) We define a subtracted correlator by $\bar{\Pi}(q^2) = \Pi(q^2) + \frac{f_\pi^2}{q^2}$ and a subtracted spectral function by
$\bar{\rho}_A(s) = \rho_A(s) - \pi f_\pi^2 \delta(s)$. Now, the $S$ parameter is given by
\beq
S = 4 \pi \bar{\Pi}(0) = 4 \int_0^\infty ds~\frac{1}{s} \left(\rho_V(s) - \bar{\rho}_A(s)\right).
\eeq
Interestingly, there are multiple well-established nonperturbative facts about $\Pi_V - \Pi_A$,
but none are sufficient to prove that $S > 0$. There are the famous Weinberg sum rules~\cite{Weinberg}
\beq
\frac{1}{\pi} \int_0^\infty ds~\left(\rho_V(s) - \bar{\rho}_A(s)\right) & = & f_\pi^2, \\
\frac{1}{\pi} \int_0^\infty ds~s\left(\rho_V(s) - \bar{\rho}_A(s)\right) & = & 0.
\eeq
Further, Witten proved that $\Sigma(Q^2)=-Q^2 (\Pi_V(Q^2) - \Pi_A(Q^2)) > 0$ for all Euclidean momenta $Q^2 = -q^2 > 0$~\cite{Witten}. However, the
positivity of $S$ seems to be more difficult to prove.

Our plausibility argument is based on the function $\Sigma(Q^2)$. In
terms of this function, $S = -4 \pi \Sigma'(0)$. (Note that in
$\Sigma(Q^2)$ the $1/Q^2$ pole from $\Pi_A$ is multiplied by $Q^2$,
yielding a constant that does not contribute when we take the
derivative. Thus when considering $\Sigma$ we do not need to
subtract the pion pole as we did in $\bar{\Pi}$.) We also know that
$\Sigma(0) = f_\pi^2 > 0$. On the other hand, we know something else
that is very general about theories that spontaneously break chiral
symmetry: at very large Euclidean $Q^2$, we should see symmetry
restoration. More specifically, we expect behavior like \beq
\Sigma(Q^2) \rightarrow {\cal O}\left(\frac{1}{Q^{2k}}\right), \eeq
where $k$ is associated with the dimension of some operator that
serves as an order parameter for the symmetry breaking. (In some 5D
models the decrease of $\Pi_A - \Pi_V$ will actually be {\em
faster}, e.g. in Higgsless models one has exponential decrease.)
While we are most familiar with this from the OPE of QCD, it should
be very general. If a theory did not have this property and $\Pi_V$
and $\Pi_A$ differed significantly in the UV, we would not view it
as a spontaneously broken symmetry, but as an explicitly broken one.
Now, in this context, positivity of $S$ is just the statement that,
because $\Sigma(Q^2)$ begins at a positive value and eventually
becomes very small, the smoothest behavior one can imagine is that
it simply decreases monotonically, and in particular, that
$\Sigma'(0) < 0$ so  that $S > 0$.\footnote{For a related discussion
of the behaviour of $\Sigma \left( Q^2 \right)$ in the case of
large-$N_c$ QCD, see~\cite{friot}.} The alternative would be that
the chiral symmetry breaking effects push $\Sigma(Q^2)$ in different
directions over different ranges of $Q^2$. We have not proved that
this is impossible in arbitrary theories, but it seems plausible
that the simpler case is true, namely that chiral symmetry
restoration always acts to decrease $\Sigma(Q^2)$ as we move to
larger $Q^2$. Indeed, we will show below that in a wide variety of
perturbative holographic theories $S$ is positive.

\section{Boundary-effective-action approach to oblique corrections. Simple cases with boundary breaking}
\label{review}
\setcounter{equation}{0}
\setcounter{footnote}{0}

In this section we review the existing results and calculational
methods for the electroweak precision observables (and in particular
the $S$-parameter) in holographic models of electroweak symmetry
breaking. There are two equivalent formalisms for calculating these
parameters. One is using the on-shell wave function of the $W$/$Z$
bosons~\cite{onshell}, and the electroweak observables are calculated from integrals
over the extra dimension involving these wave functions. The
advantage of this method is that since it uses the physical wave
functions it is easier to find connections to the $Z$ and the KK mass
scales. The alternative formalism proposed by Barbieri, Pomarol and
Rattazzi~\cite{BPR} (and later extended in~\cite{BPRS} to include
observables off the $Z$-pole) uses the method of the boundary
effective action~\cite{HolographicPrescription},
and involves off-shell wave functions of the
boundary fields extended into the bulk. This latter method leads
more directly to a general expression of the electroweak parameters,
so we will be applying this method throughout this paper. Below we
will review the basic expressions from~\cite{BPR}.

A theory of electroweak symmetry breaking with custodial symmetry
has an SU(2)$_L\times$ SU(2)$_R$ global symmetry, of which the
SU(2)$_L\times$U(1)$_Y$ subgroup is gauged (since the $S$-parameter
is unaffected by the extra $B-L$ factor we will ignore it in our
discussion). At low energies, the global symmetry is broken to
SU(2)$_D$. In the holographic picture of~\cite{BPR} the
elementary SU(2)$\times$U(1) gauge fields are extended into the bulk
of the extra dimension. The bulk wave functions are determined by
solving the bulk EOM's as a function of the boundary fields, and the
effective action is just the bulk action in terms of the boundary
fields.

In order to first keep the discussion as general as possible, we use
an arbitrary background metric over an extra dimension parametrized
by $0<y<1$, where $y=0$ corresponds to the UV boundary, and $y=1$ to
the IR boundary. In order to simplify the bulk equations of motion
it is preferential to use the coordinates in which the metric takes
the form~\footnote{In this paper, we use a $(-+\ldots+)$ signature.
5D bulk  indices are denoted by capital Latin indices while we use
Greek letters for 4D spacetime indices. 5D indices will be raised
and lowered using the 5D metric while the 4D Minkowski metric is
used for 4D indices.}~\cite{BPR}
\begin{equation}
ds^2= e^{2\sigma} dx^2+e^{4\sigma} dy^2 \ .
\label{metric}
\end{equation}
The bulk action for the gauge fields is given by
\begin{equation}
\mathcal{S}=-\frac{1}{4g_5^2} \int d^5 x \sqrt{-g} \left( (F_{MN}^L)^2+
(F_{MN}^R)^2 \right).
\label{bulkaction}
\end{equation}
The bulk equations of motion are given by
\begin{equation}
\partial_y^2 A_\mu^{L,R}-p^2 e^{2\sigma} A_\mu^{L,R}=0,
\label{bulkEOM}
\end{equation}
or equivalently the same equations for the combinations
$V_\mu,A_\mu =(A_{\mu L}\pm A_{\mu R})/\sqrt{2}$.

We assume  that the (light) SM fermions are effectively localized on the Planck brane and that they carry their usual quantum numbers under $SU(2)_L\times U(1)_Y$ that remains unbroken on the UV brane. The values of these fields on the UV brane have therefore a standard couplings to fermion and they are the 4D interpolating fields we want to compute an effective action for. This dictates the boundary conditions we want to impose on the UV brane
\begin{equation}
    \label{eq:UVBC}
A^{L\, a}_\mu (p^2,0)= \bar{A}^{L\, a}_\mu (p^2), \
A^{R\, 3}_\mu (p^2,0)= \bar{A}^{R\, 3}_\mu (p^2), \
A^{R\, 1,2}_\mu (p^2,0)= 0 .
\end{equation}
$A_R^{1,2}$ are vanishing because they correspond to ungauged symmetry generators.
The solutions of the bulk equations of motion satisfying these UV BC's take the form
\begin{equation}
V_\mu (p^2,y)= v(y,p^2) \bar{V}_\mu (p^2), \   A_\mu
(p^2,y)= a(y,p^2) \bar{A}_\mu(p^2).
\end{equation}
where the interpolating functions $v$ and $a$ satisfy the bulk equations
\begin{equation}
\partial_y^2 f (y,p^2) - p^2 e^{2 \sigma} f(y,p^2)=0
\end{equation}
and the UV BC's
\begin{equation}
v(0,p^2)=1,\ a(0,p^2)=1.
\end{equation}

The effective action for the boundary fields reduces to a pure boundary
term since by integrating by parts the bulk action vanishes by the
EOM's:
\begin{equation}
\mathcal{S}_{\textit{eff}}=\frac{1}{2g_5^2}
\int d^4x (V_\mu \partial_y V^\mu+A_\mu\partial_y A^\mu)|_{y=0}
=\frac{1}{2g_5^2} \int d^4p  ( \bar{V}_\mu^2 \partial_y v
+ \bar{A}_\mu^2 \partial_y a )|_{y=0}
\end{equation}
And we obtain the non-trivial vacuum polarizations for the boundary vector fields
\begin{equation}
\Sigma_V (p^2) = - \frac{1}{g_5^2} \partial_y v(0,p^2), \
\Sigma_A (p^2) = - \frac{1}{g_5^2} \partial_y a(0,p^2).
\end{equation}

The various oblique electroweak
parameters are then obtained from the momentum expansion of the
vacuum polarizations in the effective action,
\begin{equation}
\Sigma (p^2)=\Sigma (0)+p^2 \Sigma'(0)+\frac{p^4}{2}
\Sigma''(0)+\ldots \end{equation} For example the $S$-parameter is
given by
\begin{equation}
S =16 \pi \Sigma'_{3B}(0) = 8\pi (\Sigma'_V (0) - \Sigma'_A (0)).
\end{equation}
A similar momentum expansion can be performed on the interpolating functions $v$ and $a$:
$v(y,p^2)=v^{(0)}(y)+p^2 v^{(1)}(y)+\ldots$, and similarly for $a$.
The $S$-parameter is then simply expressed as
\begin{equation}
S=-\frac{8\pi}{g_5^2} (\partial_y v^{(1)}-\partial_y a^{(1)})|_{y=0}.
\end{equation}

The first general theorem was proved in~\cite{BPR}: for the case of
boundary condition breaking in a general metric, $S\geq 0$. The proof
uses the explicit calculation of the functions
$v^{(n)},a^{(n)}, n=0,1$.
First,  the bulk equations~(\ref{bulkEOM}) write
\begin{equation}
\partial_y^2 v^{(0)}=\partial_y^2 a^{(0)}=0, \ \partial_y^2
v^{(1)}=e^{2\sigma} v^{(0)}, \ \partial_y^2 a^{(1)}=e^{2\sigma}
a^{(0)}.
\end{equation}
And the $p^2$-expanded UV BC's are
\begin{equation}
v^{(0)}=a^{(0)}=1, v^{(1)}=a^{(1)}=0 \ \textrm{at } y=0
\end{equation}
Finally, we need to specify the BC's on the IR brane that correspond to the breaking
$SU(2)_L\times$SU(2)$_R\to$ SU(2)$_D$
\begin{equation}
        \label{eq:IRBCbreaking}
\partial_y V_\mu =0,\ A_\mu =0,
\end{equation}
which translates into simple BC's for the interpolating functions
\begin{equation}
\partial_y v^{(n)}=a^{(n)}=0,\ n=0,1.
\end{equation}
The solution of these equations are
$v^{(0)}=1, a^{(0)}=1-y,
v^{(1)}=\int_0^y dy' \int_0^{y'} dy'' e^{2\sigma(y'')} - y \int_0^1 dy' e^{2 \sigma (y')} $,
$a^{(1)}=\int_0^y dy' \int_0^{y'} dy'' e^{2\sigma (y'')}(1-y'') -y \int_0^1dy' \int_0^{y'} dy'' e^{2\sigma (y'')}(1-y'')$.
Consequently
\begin{equation}
S=\frac{8\pi}{g_5^2} \left( \int_0^1 dy e^{2 \sigma (y)}dy
-\int_0^1 dy \int_0^y dy' (1-y') e^{2 \sigma (y')} \right)
\end{equation}
which is manifestly positive.

\subsection{$S>0$ for BC breaking with boundary kinetic mixing}

The first simple generalization of the BC breaking model is to
consider the same model but with an additional localized kinetic
mixing operator added on the TeV brane (the effect of this operator has been
studied in flat space in~\cite{BPR} and in AdS space in~\cite{onshell}).
The localized Lagrangian is
\begin{equation}
- \frac{\tau}{4g_5^2} \int d^4x \sqrt{-g} V_{\mu\nu}^2  .
\end{equation}
This contains only the kinetic term for the vector field since the
axial gauge field is set to zero by the BC breaking. In this case the BC at
$y=1$ for the vector field is modified to $\partial_y V_\mu+\tau p^2
V_\mu=0$. In terms of the wave functions expanded in small momenta we
get $\partial_y v^{(1)}+\tau v^{(0)}=0$. The only change in the
solutions will be that now $v^{(1)'}= -\tau -\int_y^1 e^{2\sigma
(y')} dy'$, resulting in
\begin{equation}
S=\frac{8\pi}{g_5^2} \left( \int_0^1 e^{2 \sigma (y)}dy-\int_0^1
dy \int_0^y (1-y') e^{2 \sigma (y')} dy' +\tau \right)
\end{equation}
Thus as long as the localized kinetic term has the proper sign, the
shift in the $S$-parameter will be positive. If the sign is
negative, there will be an instability in the theory since fields
localized very close to the TeV brane will feel a wrong sign kinetic
term. Thus we conclude that for the physically relevant case $S$
remains positive.

\subsection{$S>0$ for BC breaking with arbitrary kinetic functions}
\label{sec:BCkin}

The next simple extension of the BPR result is to consider the case
when there is an arbitrary $y$-dependent function in front of the
bulk gauge kinetic terms. These could be interpreted as effects of
gluon condensates modifying the kinetic terms in the IR. In this
case the action is
\begin{equation}
\mathcal{S}=-\frac{1}{4g_5^2} \int d^5 x \sqrt{-g} \left( \phi_L^2(y)
(F_{MN}^L)^2+\phi_R^2(y) (F_{MN}^R)^2 \right).
\end{equation}
 $\phi_{L,R} (y)$ are arbitrary profiles for the
gauge kinetic terms, which are assumed to be the consequence of some
bulk scalar field coupling to the gauge fields. Note that this case
also covers the situation when the gauge couplings are constant but
$g_{5L}\neq g_{5R}$. The only assumption we are making is that the
gauge kinetic functions for $L,R$ are strictly positive. Otherwise one
could create a wave packet localized around the region where the
kinetic term is negative which would have ghost-like behavior.

Due to the $y$-dependent kinetic terms it is not very useful to go
into the $V,A$ basis. Instead we will directly solve the bulk
equations in the original basis. The bulk equations of motion for
$L,R$ are given by
\begin{equation}
\partial_y (\phi_{L,R}^2 \partial_y A_\mu^{L,R})-p^2 e^{2\sigma}
\phi_{L,R}^2A_\mu^{L,R}=0
\end{equation}
To find the boundary effective action needed to evaluate the
S-parameter we perform the following decomposition:
\begin{eqnarray}
&& A_\mu^L(p^2,y)=\bar{L}_\mu(p^2) L_L (y,p^2)+\bar{R}_\mu (p^2)L_R
(y,p^2) , \nonumber \\
&& A_\mu^R(p^2,y)=\bar{L}_\mu(p^2) R_L (y,p^2)+\bar{R}_\mu (p^2) R_R
(y,p^2) .
\end{eqnarray}
Here $\bar{L},\bar{R}$ are the boundary fields, and the fact that we
have four wave functions expresses the fact that these fields will
be mixing due to the BC's on the IR brane. The UV BC's~(\ref{eq:UVBC}) and the IR BC's~(\ref{eq:IRBCbreaking}) can be written in terms of the interpolating functions as
\begin{eqnarray}
{\rm (UV)} && L_L(0,p^2)=1, \ L_R(0,p^2)=0, \ R_L(0,p^2)=0, \
R_R(0,p^2)=1.
\\[.3cm]
{\rm (IR)} &&
\begin{array}{l}
L_L(1,p^2)=R_L(1,p^2), \ L_R(1,p^2)=R_R(1,p^2), \\[.1cm]
\partial_y (L_L(1,p^2)+R_L(1,p^2))=0, \ \partial_y (L_R(1,p^2)+R_R(1,p^2))=0.
\end{array}
\end{eqnarray}
The solution of these equations with the proper boundary conditions and for small values of $p^2$
is rather lengthy, so we have placed the details in
Appendix~\ref{app:boringdetails}. The end result is that
\begin{equation}
S=  - \frac{8 \pi}{g_5^2} \left( \phi_L^2 \partial_y L_R^{(1)} + \phi_R^2 \partial_y R_L^{(1)} \right) |_{y=0}
= - \frac{8 \pi}{g_5^2} (a_{L_R}+a_{R_L}),
\end{equation}
where the constants $a_{R_L}$ are negative as their explicit expressions shows it.
Therefore $S$ is positive.

\section{$S>0$ in models with bulk Higgs}
\label{bulkbreaking}
\setcounter{equation}{0}
\setcounter{footnote}{0}

Having shown than $S > 0$ for arbitrary metric and EWSB through BC's, in this section,
we switch to considering breaking of electroweak symmetry by a bulk scalar (Higgs) vev.
We begin by neglecting the effects of kinetic mixing
between $SU(2)_L$ and $SU(2)_R$ fields coming from higher-dimensional
operator in the $5D$ theory, expecting that their effect,
being suppressed
by the $5D$ cut-off, is sub-leading. We will return to a consideration
of such kinetic mixing effects in the following sections.

We will again use the metric (\ref{metric}) and the bulk action (\ref{bulkaction}).
Instead of BC breaking we assume that EWSB is caused by a bulk Higgs which
results in a $y$-dependent profile for the axial mass term
\begin{equation}
-\int d^5 x \sqrt{-g}\, \frac{M^2(y)}{2g_5^2} A_M^2.
\end{equation}
Here $M^2$ is a positive function of $y$ corresponding to the background Higgs
VEV. The bulk equations of motion are:
\begin{equation}
(\partial_y^2 - p^2 e^{2 \sigma})V_\mu  =  0, \ \
(\partial_y^2 - p^2 e^{2 \sigma} - M^2 e^{4 \sigma}) A_\mu  =  0.
    \label{bulkAeq}
\end{equation}
On the IR brane, we want to impose regular Neumann BC's that preserve the full SU(2)$_L \times$ SU(2)$_R$ gauge symmetry
\begin{equation}
    \label{eq:IRBCbulkbreaking}
(IR)\ \ \partial_y V_\mu = 0, \ \ \partial_y A_\mu = 0.
\end{equation}
As in the previous section, the BC's on the UV brane just define the 4D interpolating fields
\begin{equation}
    \label{eq:UVBCbulkbreaking}
(UV)\ \  V_\mu (p^2,0) =  \bar{V}_\mu (p^2), \  \ A_\mu (p^2,0) = \bar{A}_\mu (p^2).
\end{equation}
The solutions of the bulk equations of motion satisfying these BC's take the form
\begin{equation}
V_\mu (p^2,y)= v(y,p^2) \bar{V}_\mu (p^2), \  \  A_\mu
(p^2,y)= a(y,p^2) \bar{A}_\mu(p^2),
\end{equation}
where the interpolating functions $v$ and $a$ satisfy the bulk equations
\begin{equation}
    \label{eq:bulkEOMbulkbreaking}
\partial_y^2 v  - p^2 e^{2 \sigma} v=0,\ \
\partial_y^2 a  - p^2 e^{2 \sigma} a - M^2 e^{4\sigma} a=0.
\end{equation}
As before, these interpolating functions are expanded in powers of the momentum:
$v(y,p^2)=v^{(0)}(y)+p^2 v^{(1)}(y)+\ldots$, and similarly for $a$.
The $S$-parameter is again given by the same expression
\begin{equation}
S=-\frac{8\pi}{g_5^2} (\partial_y v^{(1)}-\partial_y a^{(1)})|_{y=0}.
\end{equation}
We will not be able to find general solutions for  $\aone$ and $\vone$ but we are going to
prove that $\partial_y \aone > \partial_y \vone$ on the UV brane, which is exactly what is needed to conclude that $S>0$.

First at the zeroth order in $p^2$,  the solution for $\vo$ is simply constant, $\vo=1$, as before. And $\ao$  is the solution of
\begin{equation}
\partial_y^2 \ao = M^2 e^{4 \sigma} \ao, \ \  \ao|_{y=0}=1, \ \ \partial_y \ao |_{y=1} = 0.
\end{equation}
In particular, since $\ao$ is positive at $y = 0$, this implies that $\ao$
remains positive: if $\ao$ crosses through zero it must be decreasing, but
then this equation shows that the derivative will continue to decrease and can
not become zero to satisfy the other boundary condition. Now, since $\ao$ is
positive, the equation of motion shows that it is always concave up, and then
the condition that its derivative is zero at $y = 1$ shows that it is a
decreasing function of $y$. In particular, we have for all $y$
\begin{equation}
\ao(y) \le \vo(y),
\end{equation}
with equality only at $y = 0$.

Next consider the order $p^2$ terms. What we wish to show is that  $\partial_y \aone > \partial_y \vone$ at the UV brane. First, let's examine the behavior of
$\vone$: the boundary conditions are $\vone |_{y=0} = 0$ and $\left. \partial_y
  \vone \right|_{y=1} = 0$. The equation of motion is:
\begin{equation}
\partial_y^2 \vone = e^{2\sigma} \vo = e^{2\sigma} > 0,
\end{equation}
so the derivative of $\vone$ must increase to reach zero at $y=1$. Thus it is
negative everywhere except $y=1$, and $\vone$ is a monotonically decreasing
function of $y$. Since $\vone|_{y=0}=0$, $\vone$ is strictly negative on $(0,1]$.

For the moment suppose that $\aone$ is also strictly negative; we will provide
an argument for this shortly. The equation of motion for $\aone$ is:
\begin{equation}
\partial_y^2 \aone = e^{2\sigma} \ao + M^2 e^{4\sigma} \aone.
\end{equation}
Now, we know that $\ao < \vo$, so under our assumption that $\aone < 0$, this
means that
\begin{equation}
\partial_y^2 \aone \leq \partial_y^2 \vone,
\end{equation}
with equality only at $y=0$. But we also know that $\partial_y \vone \partial_y \aone$ at $y=1$, since they both satisfy Neumann boundary
conditions there. Since the derivative of $\partial_y \aone$ is strictly
smaller over $(0,1]$, it must start out at a higher value in
order to reach the same boundary condition. Thus we have that
\begin{equation}
\left. \partial_y \aone \right|_{y=0} > \left. \partial_y \vone \right|_{y=0}.
\end{equation}

The assumption that we made is that $\aone$ is strictly negative over
the interval $(0,1]$. The reason is the following: suppose that $\aone$
becomes positive at some value of $y$. Then as it passes through zero it is
increasing. But then we also have that $\partial_y^2 \aone = e^{2\sigma} \ao +
M^2 e^{4\sigma} \aone$, and we have argued above that $\ao > 0$. Thus if
$\aone$ is positive, $\partial_y \aone$ remains positive, because
$\partial_y^2 \aone$ cannot become negative. In particular, it becomes
impossible to reach the boundary condition $\partial_y \aone = 0$ at
$y=1$. This fills the missing step in our argument and shows that the $S$
parameter must be positive.

In the rest of this section we show that the above proof for the
positivity of S remains
essentially unchanged in the case when the bulk gauge couplings for
the SU(2)$_L$ and SU(2)$_R$ gauge groups are not equal. In this case
(in order to get diagonal bulk equations of motion) one needs to
also introduce the canonically normalized gauge fields. We start
with the generic action (metric factors are understood when
contracting indices)
\begin{equation}
\int d^5 x \sqrt{-g} \left( -\frac{1}{4 g_{5L}^2}
(F_{MN}^L)^2-\frac{1}{4 g_{5R}^2} (F_{MN}^R)^2-\frac{h^2(z)}{2}
(L_M -R_M )^2 \right)
\end{equation}
To get to a canonically normalized diagonal basis we redefine the
fields as
\begin{equation}
\tilde{A} = \frac{1}{\sqrt{g_{5L}^2+g_{5R}^2}} (L-R)\ , \ \tilde{V}
= \frac{1}{\sqrt{g_{5L}^2+g_{5R}^2}}
\left(\frac{g_{5R}}{g_{5L}}L+\frac{g_{5L}}{g_{5R}} R\right)\ .
\end{equation}
To get the boundary effective action, we write the fields
$\tilde{V}, \tilde{A}$ as
\beq
\tilde{A}(p^2,z) & = &\frac{1}{\sqrt{g_{5L}^2+g_{5R}^2}}
\left(\bar{L}(p^2)-\bar{R}(p^2)\right) \tilde{a}(p^2,z) \ , \ \\
\tilde{V}(p^2,z) & = & \frac{1}{\sqrt{g_{5L}^2+g_{5R}^2}}
\left(\frac{g_{5R}}{g_{5L}}\bar{L}(p^2)+\frac{g_{5L}}{g_{5R}}
\bar{R}(p^2)\right)\tilde{v}(p^2,z)\ .
\eeq

Here $\bar{L}, \bar{R}$ are the boundary effective fields (with
non-canonical normalization exactly as in~\cite{BPR}), while the
profiles $\tilde{a}, \tilde{v}$ satisfy the same bulk equations and
boundary conditions as $a,v$ in (\ref{bulkAeq})--(\ref{eq:UVBCbulkbreaking}) with
an appropriate replacement for $M^2=(g_{5L}^2+g_{5R}^2) h^2$. In terms of the canonically normalized fields, the boundary effective action takes its usual form
\begin{equation}
\mathcal{S}_{\textit{eff}}=
\frac{1}{2} \int d^4x \left( \tilde{V} \partial_y \tilde{V} +\tilde{A}
\partial_y \tilde{A} \right)_{y=0}.
\end{equation}
And we deduce the vacuum polarization
\begin{equation}
\Sigma_{L3 B} (p^2) =
- \frac{1}{g_{5L}^2+g_{5R}^2} (\partial_y
\tilde{v}(0,p^2)- \partial_y \tilde{a}(0,p^2))
\end{equation}
And finally the $S$-parameter is equal to
\begin{equation}
S= - \frac{16 \pi }{g_{5L}^2 + g_{5R}^2} ( \partial_y \tilde{v}^{(1)} - \partial_y \tilde{a}^{(1)} )
\end{equation}
Since $\tilde{a}^{(n)},\tilde{v}^{(n)}, n=0,1$ satisfy the same equations (\ref{bulkAeq})--(\ref{eq:UVBCbulkbreaking}) as before,  the proof goes through unchanged and we conclude that $S>0$.

\section{Bulk Higgs and bulk kinetic mixing}
\label{equivalence}
\setcounter{equation}{0}
\setcounter{footnote}{0}

Next, we wish to consider the effects of kinetic mixing from
higher-dimensional operator in the bulk involving the Higgs VEV -- as mentioned earlier,
this kinetic mixing is suppressed
by the $5D$ cut-off and hence expected to be
a sub-leading effect.
The reader
might wonder
why we neglected it before, but consider it now? The point
is that,
although the
leading effect on $S$ parameter is positive as shown above, it can be
accidentally suppressed so that the {\em formally} sub-leading effects
from the bulk kinetic mixing can be
important, in particular, such effects could change the sign of $S$.
Also,
the Higgs
VEV can be large, especially
when the Higgs profile is ``narrow'' such that it approximates BC breaking,
and thus the large VEV can (at least partially)
compensate the suppression from the $5D$
cut-off.
Of course, in this limit of BC breaking ($\delta$-function
VEV), we know that kinetic mixing gives $S < 0$ only if tachyons
are present in the spectrum,
but we would like to cover the cases intermediate
between BC breaking limit and a broad Higgs profile as well.
In this section, we develop a formalism, valid for
arbitrary metric and Higgs profile, to treat the bulk mass term
and kinetic mixing on the same footing and then we apply
this technique to models in AdS space
and with power-law profiles for Higgs VEV in the next section.

We first present a discussion of how a profile
for the $y$-dependent kinetic term is equivalent to a bulk mass term. This is equivalent to the result~\cite{HirnSanz} that a bulk mass term can be equivalent to an effective metric. However, we find the particular formulation that we present here to be more useful when we deal with the case of a kinetic mixing. Assume we have a Lagrangian for a gauge field that has a kinetic
term
\begin{equation}
\mathcal{S}=-\frac{1}{4g_5^2} \int d^5 x \sqrt{-g} \phi^2 (y) F_{MN}^2
\end{equation}
We work in the axial gauge $A_5=0$ and again the metric takes the form (\ref{metric}).
We redefine the field to absorb the function $\phi$:
$\tilde{A}(y)=\phi (y) A(y)$. The action in terms of the new field
is then written as
\begin{equation}
\mathcal{S}= -\frac{1}{4g_5^2} \int d^5 x \left( e^{2\sigma} \tilde{F}_{\mu\nu}^2
+2 (\partial_y \tilde{A}_\mu)^2
+2 \frac{\phi'^2}{\phi^2}\tilde{A}_\mu^2 -4 (\partial_y \tilde{A}_\mu)
\tilde{A}^\mu \frac{\phi'}{\phi}\right)
\end{equation}
To see that the kinetic profile $\phi$ is equivalent to a mass term, we
integrate by parts in the second term
\begin{equation}
\mathcal{S}=-\frac{1}{4g_5^2} \int d^5 x \sqrt{-g} \left( \tilde{F}_{MN}^2 +2
e^{-4\sigma} \frac{\phi''}{\phi} \tilde{A}_\mu^2\right)
+\frac{1}{2g_5^2} \int d^4 x \left. \frac{\phi'}{\phi} \tilde{A}_\mu^2 \right|_0^1
\end{equation}
Thus we find that a bulk kinetic profile is equivalent to a bulk
mass plus a boundary mass. The bulk equations of motion for the new variables
will then be
\begin{equation}
\partial_y^2 {\tilde A}_\mu - e^{2\sigma} p^2 {\tilde A}_\mu  -\frac{\phi^{\prime\prime}}{\phi} {\tilde A}_\mu =0,
\end{equation}
and the boundary conditions become
\begin{equation}
\partial_y {\tilde A}_\mu = \frac{\phi^\prime}{\phi} {\tilde A}_\mu.
\end{equation}
Note, that despite the bulk mass term, there is still a massless mode whose wavefunction is simply $\phi (z)$.
Now we can reverse the argument and say that a bulk mass must be
equivalent to a profile for the bulk kinetic term plus a boundary mass term.

\subsection{The general case}
%
%

We have seen above how to go between a bulk mass terms and a kinetic
function. We will now use this method to discuss the general case, when
there is electroweak symmetry breaking due to a bulk higgs with a sharply
peaked profile toward the IR brane, and the same Higgs introduces kinetic mixing
between L and R fields corresponding to a higher dimensional operator
from the bulk. For now we assume
that the Higgs fields that breaks the electroweak symmetry is in a
(2,2) of SU(2)$_L\times$SU(2)$_R$, with a VEV $\langle H \rangle
={\rm diag} (h(z),h(z))/\sqrt{2}$.\footnote{An alternative possibility would be to consider a Higgs in the
(3,3) representation of SU(2)$_L\times$SU(2)$_R$.} This Higgs profile $h$ has
dimension 3/2.  The 5D action is given by
\begin{equation}
\int d^5 x \sqrt{-g} \left[ -\frac{1}{4g_5^2} \left[
(F_{MN}^L)^2+(F_{MN}^R)^2\right] - (D_MH)^\dagger
(D^MH)+\frac{\alpha}{\Lambda^2} {\rm Tr} (F_{MN}^L H^\dagger H
F^{MN\ R})\right].
\end{equation}
Here $\alpha$ is a coefficient of ${\cal O}(1)$ and $\Lambda$ is the
5D cutoff scale, given approximately by $\Lambda \sim
24 \pi^3/ g_5^2$. The kinetic mixing term just generates a
shift in the kinetic terms of the vector and axial vector field, and
we will write the bulk mass term also as a shift in the kinetic term
for the axial vector field. The exact form of the translation
between the two forms is given by answering the question of how to
redefine the field with an action (note that $m^2$ has a mass dimension 3)
\begin{equation}
-\frac{1}{4g_5^2} \int d^5 x \sqrt{-g} \left( w F_{MN}^2 + m^2 2
g_5^2 A_\mu A^\mu \right)
\end{equation}
to a theory with only a modified kinetic term. The appropriate field
redefinition $A=\rho \tilde{A}$ will be canceling the mass term if
$\rho$ satisfies
\begin{equation} \label{eq:rhoeom}
\partial_y (w\partial_y \rho ) = m^2 g_5^2 e^{4 \sigma} \rho  ,
\end{equation}
together with the boundary conditions $\rho'|_{y=1}=0, \rho|_{y=0}=1$. The
relation between the new and the old expression for $w$ will be
$\tilde{w}= \rho^2 w$. The action in this case is given by
\begin{equation}
-\frac{1}{4g_5^2} \int d^5 x \sqrt{-g} \tilde{w} \tilde{F}_{MN}^2
+\int d^4 x \frac{\tilde{w}(0)}{2 g_5^2} (\partial_y \rho )\tilde{A}^2|_{y=0}
\end{equation}
This last boundary term is actually irrelevant for the $S$-parameter:
since it does not contain a derivative on the field it can not get
an explicit $p$-dependence so it will not contribute to $S$, so for
practical purposes this boundary term can be neglected.

With this expression we now can calculate $S$. For this we need the
modified version of the formula from~\cite{HirnSanz}, where the breaking is not
by boundary conditions but by a bulk Higgs. The expression is
\begin{equation}
S=\frac{8\pi}{g_5^2}\int_{0}^{1} dy e^{2\sigma} (w_V- {\tilde w}_A).
\label{Sexpr}
\end{equation}
 In our case $w_V=1-\frac{ \alpha h^2 (y)  2 g_5^2}{\Lambda^2}$ while
${\tilde w}_A=w_A \rho^2=(1+\frac{ \alpha h^2 (y) 2 g_5^2}{\Lambda^2})\rho^2$.

This formula also gives another way to see that $S > 0$ in the absence of kinetic mixing,
without analyzing the functions $v^{(1)}$ and $a^{(1)}$ from Section~\ref{bulkbreaking}
in detail. Without kinetic mixing, $w_V = 1$ and ${\tilde w}_A = \rho^2$,
and the equation of motion for $\rho$ is simply $\partial_y^2 \rho = m^2 g_5^2  e^{4\sigma}\rho$.
In that case $\rho$ is just the function we called $a^{(0)}$ in Section~\ref{bulkbreaking}.
Since we showed there that $a^{(0)} \leq 1$, we see that our expression~\ref{Sexpr}
gives an alternative argument that $S > 0$ without kinetic mixing, because it is simply
an integral of $e^{2\sigma} (1 - \rho^2) \geq 0$.

\subsection{Scan of the parameter space for AdS backgrounds}
\label{scan}

Having developed the formalism for a unified
treatment of bulk mass terms and bulk kinetic mixing, we then
apply it to the
AdS case with a power-law profile for the
Higgs vev. Requiring (i) calculability of the
5D theory, i.e., NDA size of the higher-dimensional operator,
(ii) that excited $W/Z$'s are heavier
than a few 100 GeV, and
(iii) a ghost-free theory, i.e., positive kinetic terms
for both $V$ and $A$ fields, we find that $S$ is always positive
in our scan for this model.
We do not have a general proof that $S > 0$ for an arbitrary
background with arbitrary Higgs profiles, if we include the effects
of the bulk kinetic mixing, but we feel that such a possibility is
quite unlikely
based on our exhaustive scan. For
this scan we will take the parametrization of the Higgs profile
from~\cite{gaugephobic}. Here the metric is taken as AdS space \beq
ds^2=  \left( \frac{R}{z}\right)^2   \Big( \eta_{\mu \nu} dx^\mu
dx^\nu - dz^2 \Big)~, \eeq where as usual $R<z<R'$. The bulk Higgs
VEV is assumed to be a pure monomial in $z$ (rather than a
combination of an increasing and a decreasing function). The reason
for this is that we are only interested in the effect of the strong
dynamics on the electroweak precision parameters.
A term in the
Higgs VEV growing toward the UV brane would mean that
the value of bulk Higgs field evaluated on the UV brane gets
a VEV, implying that
there is EWSB also by
a elementary Higgs (in addition
to the strong dynamics) in the $4D$ dual.
We do not want to
consider such
a case.
The form of the Higgs VEV is then assumed to be \beq v (z) =
\sqrt{\frac{2 (1+\beta) \log R'/R}{(1-(R/R')^{2+2\beta})}}\, \frac{g
V}{g_5} \frac{R'}{R} \left( \frac{z}{R'} \right)^{2+\beta}\,, \eeq
where the parameter $\beta$ characterizes how peaked the Higgs
profile is toward the TeV brane ($\beta \to -1$ corresponds to a
flat profile, $\beta \to \infty$ to an infinitely peaked one). The
other parameter $V$ corresponds to an ``effective Higgs VEV'', and
is normalized such that for $V\to 246$ GeV we recover the SM and the
KK modes decouple ($R'\to \infty$ irrespective of $\beta$). For more
details about the definitions of these parameters
see~\cite{gaugephobic}.\footnote{Refs.~\cite{others} also considered
similar models.}

We first numerically fix the $R'$ parameter
for every given $V,\beta $ and kinetic mixing parameter $\alpha$ by
requiring that the $W$-mass is reproduced. We do this approximately,
since we assume the simple matching relation $1/g^2=R \log
(R'/R)/g_5^2$ to numerically fix the value of $g_5$, which is only
true to leading order, but due to wave function distortions and the
extra kinetic term will get corrected.
Then, $\rho$ can be numerically calculated by solving
(\ref{eq:rhoeom}), and from this $S$ can be obtained via
(\ref{Sexpr}).

\begin{figure}
\label{fig:Scont}
\centerline{\includegraphics[width=.3\hsize]{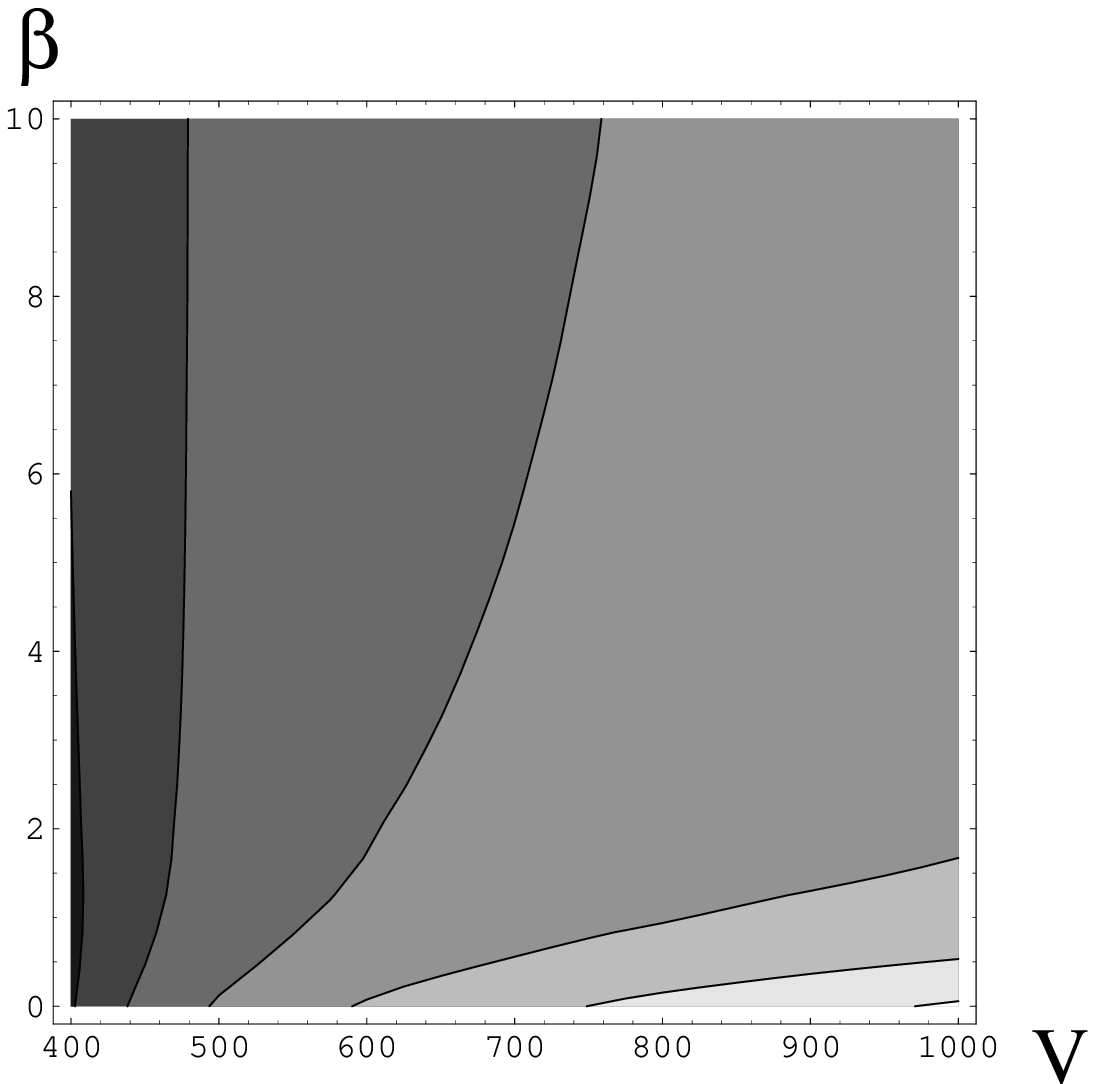}\hspace*{2cm}\includegraphics[width=.3\hsize]{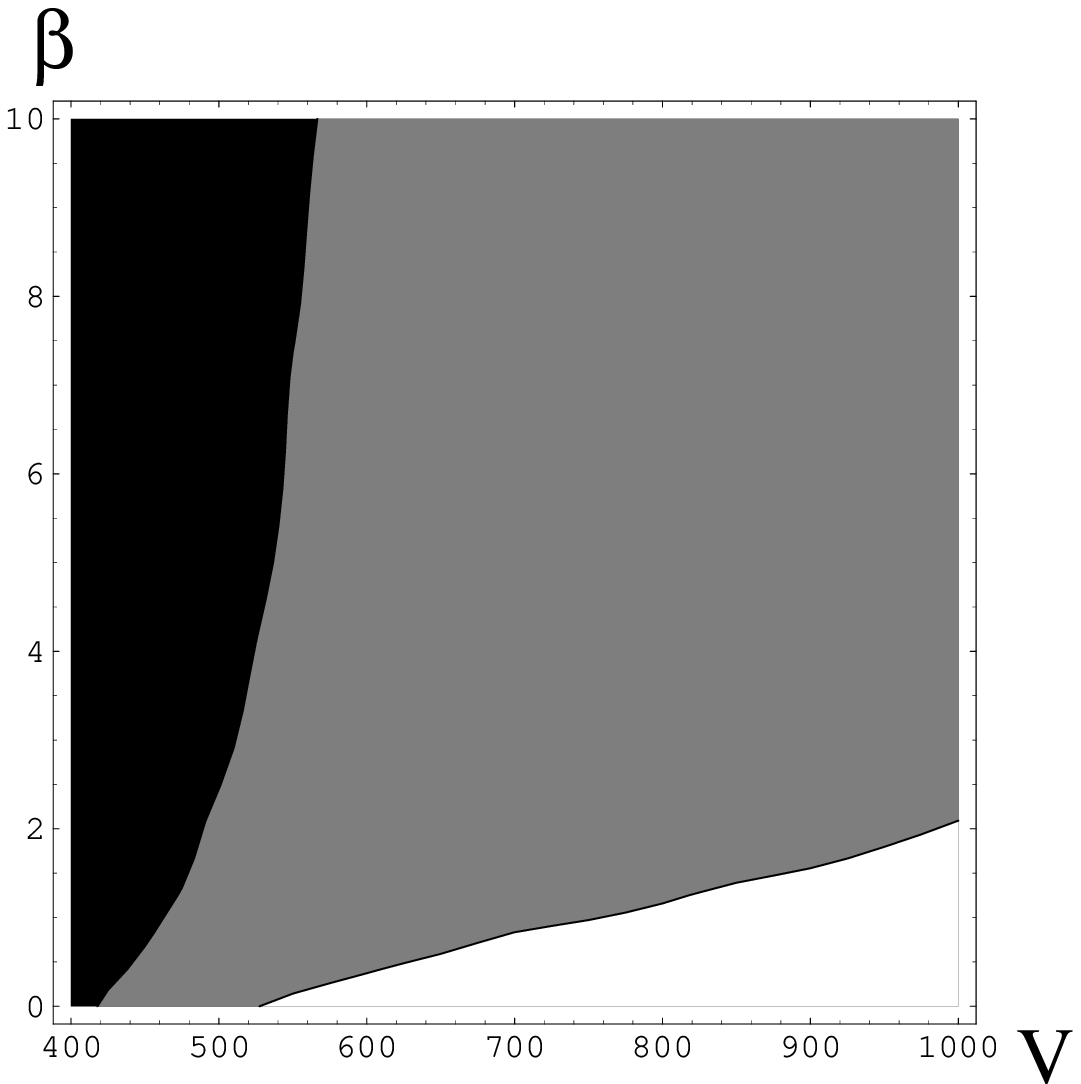}}
\caption{The contours of models with fixed values of the
$S$-parameter due to the electroweak breaking sector. In the left
panel we fix $1/R=10^8$ GeV, while in the right $1/R=10^{18}$ GeV.
The gauge kinetic mixing parameter $\alpha$ is fixed to be the
maximal value corresponding to the given $V,\beta$ (and $R'$ chosen
such that the W mass is approximately reproduced). In the left
panel the contours are $S=1,2,3,4,5,6$, while in the right
$S=1,1.5,2$.}
\end{figure}

We see that $S$ decreases as we increase $\alpha$. On the the hand,
the kinetic function for vector field ($w_V)$ also decreases
in this limit. So, in order to find the minimal value of $S$
consistent with the absence of ghosts in the theory,
we find numerically the maximal value of $\alpha$ for
every value of $V,\beta$ for which the kinetic function of the
vectorlike gauge field is still strictly positive.  We then show
contour plots for the minimal value of $S$ taking this optimal value of $\alpha$ as a
function of $V,\beta$ in Fig.~\ref{fig:Scont}. In the first figure
we fix $R'=10^{-8}$ GeV$^{-1}$, which is the usual choice for
higgsless models with light KK W' and Z' states capable of rendering
the model perturbative. In the second plot we choose the more
conventional value $R=10^{-18}$ GeV$^{-1}$. We can see the $S$ is
positive in both cases over all of the physical parameter space.

We can estimate the corrections to the above matching relation from the
wavefunction distortion and kinetic mixing as follows.
The effect
from wavefunction distortion
is expected to be $\sim
%
%
g^2 S / (16 \pi )$
which is $\stackrel{<}{\sim}
10 \%$ if we restrict to regions of parameter space with
$S \stackrel{<}{\sim} 10$.
Similarly, we estimate the effect
due to
kinetic mixing by simply
integrating the operator
over the extra dimension to find a deviation $\sim g^6 ( V R^{ \prime } )^2 \log ^2 \left(
R^{ \prime } / R \right ) / \left( 24 \pi^3 \right)^2$. So, if
restrict to $V  \stackrel{<}{\sim} 1$ TeV
and $1 / R^{ \prime } \stackrel{>}{\sim} 100$ GeV, then this deviation
is also small enough.
We see that both effects are small due to the deviation
being non-zero only near IR brane --
even though it is O(1) in
that region, whereas the
zero-mode profile used in the matching
relation is spread throughout the extra dimension.

In order to be able to make a more general statement (and to check that the
neglected additional contributions to the gauge coupling matching from
the wave function distortions and the kinetic mixing indeed do not significantly
our results) we have performed an additional scan over AdS space models where
we do not require the correct physical value of $M_W$ to be reproduced.
In this scan we then treat $R'$ as an independent free parameter. In this
case the correct matching between $g$ and $g_5$ is no longer important for
the sign of $S$, since at every place where $g_5$ appears it is multiplied
by a parameter we are scanning over anyway ($V$ or $\alpha$).

We performed
the scan again for two values of the AdS curvature, $1/R=10^{8}$ and $10^{18}$
GeV. For the first case we find that if we restrict $\alpha <10, 1/R'<1$ TeV
there is no case with $S<0$. However, there are some cases with $S<0$
for $\alpha >10$, although in these cases the theory is likely not predictive.
For $1/R=10^{18}$ GeV we find that $S<0$ only for $V\sim 250$ GeV
and $\beta \sim 0, 1/R' \sim 1$ TeV. In this case $\alpha$ is of order one
(for example $\alpha \sim 5$). This case corresponds to the composite
Higgs model of~\cite{CompositeHiggs} and it is quite plausible that at tree-level
$S<0$ if a large kinetic mixing is added in the bulk.
However in this
case EWSB is mostly due to a Higgs, albeit a composite particle
of the strong dynamics, rather than
{\em directly} by the strong dynamics,
so it does not contradict the expectation that when EWSB is triggered
directly via strong dynamics, then S is always
large and positive. However, it shows that any general proof for
$S>0$
purely based on analyzing the properties of Eqs.
(\ref{eq:rhoeom})-(\ref{Sexpr}) is doomed to
failure, since these equations contain physical situations where
EWSB is not due to the strong dynamics but due to a
%
light
Higgs in the spectrum.
%
%
Thus any general proof likely needs to include more physical requirements
on the decoupling of the physical Higgs.

\section{Conclusions}
\label{end}

In this paper, we have studied the $S$ parameter in
holographic technicolor models, focusing especially on its sign.
The motivation for our study was as follows. An alternative (to SUSY) solution to the
Planck-weak hierarchy
involves a strongly interacting
$4D$ sector spontaneously breaking the EW symmetry.
One possibility for such a strong sector is a scaled-up version of QCD as in
the traditional technicolor models. In such models, we can use the QCD data
to ``calculate'' $S$ finding $S \sim + O(1)$ which is ruled out by the
electroweak precision data.
Faced by this constraint, the idea of a
``walking'' dynamics
was proposed and it can be then argued that
$S<0$ is possible which is much
less constrained by the data, but the $S$ parameter cannot be calculated in
such models.
In short, there is a dearth of calculable models of
(non-supersymmetric) strong dynamics in $4D$.

Based on the AdS/CFT duality, the conjecture is that
certain kinds of theories of strong dynamics in $4D$ are dual
to physics of extra dimensions.
The idea then is to construct models of EWSB in an extra dimension.
Such constructions allow
more possibilities for model-building, at the same time maintaining
calculability
%
%
if
the $5D$ strong coupling scale is larger than the compactification
scale,
corresponding to large number of technicolors in the $4D$ dual.

It was
already shown that $S > 0$ for boundary condition breaking
for
arbitrary metric
(a proof
for $S > 0$ for the case of breaking by a {\em localized} Higgs vev was recently
studied in reference \cite{Delgado:2007ne}).
In this paper, we have extended
the proof for boundary condition breaking to the case
of arbitrary bulk kinetic
%
%
functions for gauge fields or gauge kinetic mixing.

Throughout this paper, we have assumed that the (light) SM fermions are effectively
localized near the UV brane so
that flavor violation
due to higher-dimensional operators in the $5D$ theory can be suppressed,
at the same time allowing for a solution to the flavor hierarchy.
Such a localization of the light SM fermions in the extra dimension is
dual to SM fermions
being ``elementary'', i.e., not mixing with composites from
the $4D$ strong sector. It is known that the $S$
parameter
can be suppressed (or even switch sign) for a flat profile for SM fermions
(or near the TeV brane) -- corresponding to mixing
of elementary fermions with composites in the $4D$ dual,
but in such a scenario flavor issues could be a problem.

We also considered the case of bulk breaking
of the EW symmetry motivated by
recent arguments that $S<0$ is possible with
different effective metrics for vector and axial fields.
For arbitrary metric and Higgs profile, we showed that
$S > 0$ at leading order, i.e.,
neglecting effects from all higher-dimensional
operators in the $5D$ theory (especially bulk kinetic mixing),
which are expected to be sub-leading
effects being suppressed by the cut-off of the $5D$ theory.
We also note that boundary mass terms can generally be mimicked
to arbitrary precision by localized contributions to the bulk scalar profile, so we do
not expect a more general analysis of boundary plus
bulk breaking to find new features.
Obtaining $S<0$ must then require either an unphysical Higgs profile or
higher-dimensional operators to contribute effects
larger than NDA size, in which case we lose calculability
of the $5D$ theory.

To make our case for $S > 0$  stronger,
we then
explored effects of the bulk kinetic mixing between
$SU(2)_{ L , R }$ gauge fields
due to Higgs vev coming from a higher-dimensional operator
in the $5D$ theory. Even though, as mentioned above,
this effect is expected to be sub-leading,
it can nevertheless
be important (especially
for the sign of $S$) if the
leading contribution to $S$ is accidentally suppressed.
Also, the large Higgs VEV, allowed for narrow profiles in the extra dimension
(approaching the BC breaking limit), can compensate the suppression due to the cut-off
in this operator.
For this analysis,
we found it
convenient to
convert bulk (mass)$^2$ for gauge fields also to kinetic functions.
Although
a general proof for $S>0$ is lacking in such a scenario,
using the above method of treating the bulk mass for axial fields,
we found that $S \sim + O(1)$
for AdS$_5$ model with power-law Higgs profile
in the viable (ghost-free) and calculable regions of the
parameter space.

In summary, our results combined
with the
%
%
previous literature strongly
suggests that $S$ is positive for {\em calculable} models
of technicolor in $4D$ and $5D$. We also presented a
plausibility argument for $S > 0$ which is valid in general, i.e.,
even for non-calculable models.

\setcounter{equation}{0}
\setcounter{footnote}{0}

\section{Acknowledgments}
\setcounter{equation}{0}
\setcounter{footnote}{0}

We thank
Giacomo Cacciapaglia, C\'edric Delaunay, Antonio Delgado, Guido Marandella, Riccardo Rattazzi,
Matthew Schwartz and Raman Sundrum for discussions.
We also thank Johannes Hirn and Veronica Sanz for comments on the manuscript.
As we were finishing the paper, we learned that
Raman Sundrum and Tom Kramer have also obtained results
similar to ours~\cite{Sundrumtoappear}.
C.C. thanks the
theory group members at CERN for their hospitality during his visit.
K.A. is supported in part by the U. S. DOE under
Contract no. DE-FG-02-85ER 40231. The research of C.C. is supported in part by the DOE OJI grant DE-FG02-01ER41206 and in part
by the NSF grant PHY-0355005. The research of M.R. is supported in part by an NSF graduate student fellowship and in part
by the NSF grant PHY-0355005.
C.G. is supported in part by the RTN European Program MRTN-CT-2004-503369 and by the CNRS/USA exchange grant 3503.

\section*{Note added}
\setcounter{equation}{0} \setcounter{footnote}{0}
After submitting our paper to the arXiv, we learned of
~\cite{Hong:2006si} which gives a proof for $S > 0$ for an arbitrary
bulk Higgs profile in AdS background that is similar to our proof in
Section 4. However, our proof of $S > 0$ is valid for a general
metric and we have also included the effect of kinetic mixing
between $SU(2)_L$ and $SU(2)_R$ fields via higher-dimensional
operator (with Higgs vev) for the calculation of $S$ in AdS
background. We thank Deog-Ki Hong and Ho-Ung Yee for pointing out
their paper to us.

\appendix

\section{Details of BC breaking with arbitrary kinetic functions}
\label{app:boringdetails}
\setcounter{equation}{0}
\setcounter{footnote}{0}

Here we present the detailed calculation of $S$ in the case with boundary breaking
and arbitrary kinetic functions described in Section~\ref{sec:BCkin}.
Recall that we had the following decomposition:
\begin{eqnarray}
&& A_\mu^L(p^2,y)=\bar{L}_\mu(p^2) L_L (y,p^2)+\bar{R}_\mu (p^2)L_R
(y,p^2) , \nonumber \\
&& A_\mu^R(p^2,y)=\bar{L}_\mu(p^2) R_L (y,p^2)+\bar{R}_\mu (p^2) R_R
(y,p^2),
\end{eqnarray}
with boundary conditions
\begin{eqnarray}
{\rm (UV)} && L_L(0,p^2)=1, \ L_R(0,p^2)=0, \ R_L(0,p^2)=0, \
R_R(0,p^2)=1.
\\[.3cm]
{\rm (IR)} &&
\begin{array}{l}
L_L(1,p^2)=R_L(1,p^2), \ L_R(1,p^2)=R_R(1,p^2), \\[.1cm]
\partial_y (L_L(1,p^2)+R_L(1,p^2))=0, \ \partial_y (L_R(1,p^2)+R_R(1,p^2))=0.
\end{array}
\end{eqnarray}
The action again reduces to a boundary term
\begin{equation}
\mathcal{S}_{\textit{eff}}= \frac{1}{2 g_5^2} \left( \phi_L^2(0) L_\mu \partial
L^\mu +\phi_R^2(0) R_\mu \partial R^\mu \right) ,
\end{equation}
so we find that
\begin{equation}
S= - \frac{8 \pi}{g_5^2} \left( \phi_L^2 \partial_y L_R^{(1)} + \phi_R^2 \partial_y R_L^{(1)} \right) |_{y=0}
\end{equation}
where we have done an expansion in terms of the momentum for all the
wave functions as usual as $L_L(y,p^2) =L_L^{(0)}(y)+p^2 L_L^{(1)}(y)
+\ldots$. The lowest order wave functions satisfy the following bulk
equations:
\begin{equation}
\partial_y(\phi_{I}^2 \partial_y I_J^{(0)})=0,
\end{equation}
where $I$ and $J$ can refer to $L$ or $R$.
Imposing the BC's these equations can be simply solved in
terms of the integrals
\begin{equation}
f_{L}(y) = \frac{\int_0^y \frac{dy'}{\phi_L^2(y')}}{\int_0^1
\frac{dy'}{\phi_L^2(y')}+\frac{\phi_R^2(1)}{\phi_L^2(1)}\int_0^1
\frac{dy'}{\phi_R^2(y')}}, \
 f_{R}(y) = \frac{\int_0^y \frac{dy'}{\phi_R^2(y')}}{\int_0^1
\frac{dy'}{\phi_R^2(y')}+\frac{\phi_L^2(1)}{\phi_R^2(1)}\int_0^1
\frac{dy'}{\phi_L^2(y')}}
\end{equation}
as
\begin{eqnarray}
&& L_L^{(0)}=1-f_L(y), \ L_R^{(0)}=f_L(y), \ R_L^{(0)}= f_R(y), \
R_R^{(0)}= 1-f_R(y). \label{lowestsols}
\end{eqnarray}
In order to actually find S we need to go one step further, that is
calculate the next order terms in the wave functions $I_J^{(1)}$.
These will satisfy the equations
\begin{equation}
\partial_y(\phi_{I}^2 \partial_y I_J^{(1)}) =e^{2\sigma}
\phi_{I}^2 I_J^{(0)},
\end{equation}
where for the $I_J^{(0)}$ we use the solutions in (\ref{lowestsols}).
The form of the solutions will be given by
\begin{equation}
I_J^{(1)} (y)=\int_0^y \frac{dy'}{\phi_{I}^2(y')}\int_0^{y'} du
e^{2\sigma} \phi_{I}^2(u) I_J^{(0)}(u) +a_{I_J} \int_0^y
\frac{dy'}{\phi_{I}^2},
\end{equation}
where $a_{I_J}$ are constants. In terms of these quantities the S-parameter is
just given by
\begin{equation}
S= - \frac{8 \pi}{g_5^2} (a_{L_R}+a_{R_L})
\end{equation}
One can again solve the boundary conditions to find the constants
$a_{R_L},a_{L_R}$. These turn out to be
\begin{eqnarray}
&& a_{R_L}
=-\frac{1}{N_R} \left[ \int_0^1 \frac{dy}{\phi_L^2(y)}
 \int_y^1 dy'e^{2\sigma(y')} \phi_L^2(y') (1-f_L(y'))
 \right. \nonumber \\
&& \left. +\frac{\phi_L^2(1)}{\phi_R^2(1)} \int_0^1 dy
e^{2\sigma(y)}\phi_R^2(y) f_R(y) \int_0^1
\frac{dy}{\phi_L^2(y)}+\int_0^1\frac{dy}{\phi_R^2(y)}\int_0^y dy'
e^{2\sigma(y')} \phi_R^2(y') f_R(y')  \right]\hspace{1.5cm}
\end{eqnarray}
where
\begin{equation}
N_R=\int_0^1\frac{dy}{\phi_R^2(y)}+\frac{\phi_L^2(1)}{\phi_R^2(1)}
\int_0^1 \frac{1}{\phi_L^2(y)}.
\end{equation}
A similar expressions applies for $a_{L_R}$ with $L\leftrightarrow R$
everywhere. Since $0<f_{L,R}<1$, we can see that every term in the
expression is manifestly positive, so $S$ is definitely positive.

\end{document}